\documentclass[12pt,titlepage]{article}

\usepackage[margin=1in]{geometry}
\usepackage{natbib}
\usepackage{url}
\usepackage{authblk}
\usepackage{setspace}
\usepackage[switch]{lineno}
\usepackage[colorlinks=true,
  linkcolor=black,
  citecolor=blue,
  urlcolor=blue]{hyperref}

\title{Generative AI in Publishing: An Editors' Panel on Ethics and
Policies}

\newif\ifblind
\ifdefined\BlindSubmission
  \blindtrue
\else
  \blindfalse
\fi

\ifblind
\author{}
\else
\author[1]{Jun Yan}
\author[2]{Mark Glickman}
\author[3]{Paul S. Albert}
\author[4]{Yulin Hswen}
\author[5]{Lan Wang}
\author[6]{Minge Xie}

\affil[1]{Department of Statistics, University of Connecticut}
\affil[2]{Department of Statistics, Harvard University}
\affil[3]{Biostatistics Branch, Division of Cancer Epidemiology and
Genetics, National Cancer Institute, National Institutes of Health}
\affil[4]{Department of Epidemiology and Biostatistics, School of
Public Health, University of Maryland}
\affil[5]{Department of Management Science, Miami Herbert Business
School, University of Miami}
\affil[6]{Department of Statistics, Rutgers, The State University of
New Jersey}
\fi

\date{}

\begin{document}

\doublespacing

\maketitle


\begin{abstract}
\noindent
Generative artificial intelligence (GenAI) is reshaping scholarly
research faster than journals have developed stable norms for its
use. This article presents an edited thematic account of a 2026 Joint
Statistical Meetings panel that brought together editorial
perspectives from mathematical statistics, data science, biomedical
statistics, and general statistical scholarship. The discussion
examines journal policies, disclosure, authorship and research
integrity, peer-review confidentiality, researcher training,
editorial workload, access, and possible future models of scholarly
publishing. Panelists shared commitments to human accountability, the
protection of confidential submissions, and disclosure of
consequential assistance, while offering different recommendations on
assistance with research ideas and proofs, disclosure requirements,
automated review, and policy enforcement. By distinguishing shared
principles from unresolved implementation questions, the article
clarifies the choices facing statistical publishing and outlines an
agenda for evaluating policies and practices as GenAI evolves. The
account seeks to foster continued discussion of GenAI in scientific
communication and encourage statistical organizations to develop more
robust operational standards.

\end{abstract}

\section{Introduction}

Generative artificial intelligence (GenAI) has already become part of
the ordinary infrastructure of scholarly work before journals,
universities, and professional societies have reached stable norms
for its use. GenAI tools, which are readily accessible to many
researchers, can improve grammar, clarify technical prose, summarize
literature, write code, propose simulations, generate images, draft
responses to reviewers, and produce text that resembles a complete
research article. These uses differ substantially in their relation
to authorship, research design, evidence, and accountability, but
they often enter a manuscript through the same interface and may be
difficult to separate after several rounds of revision. GenAI
therefore affects how authors conduct and communicate research, how
reviewers assess confidential submissions, and how editors evaluate
the integrity of the work presented to them. Recent discussions of
GenAI in science and scholarly publishing therefore highlight both
its potential to expand research capabilities and the risks it poses
to the scholarly record \citep{stokelWalkerVanNoorden2023,
bocanegraValle2026,marabelliEtAl2026}.

Developing effective policies is challenging because GenAI assistance
ranges from routine editing to contributions that shape a paper's
intellectual substance. Empirical work suggests that researchers
already use GenAI across multiple stages of the research process, so
policies that assume occasional or easily isolated use may be too
narrow for current practice \citep{andersenEtAl2025}. At the same
time, ethics discussions emphasize that human researchers remain
responsible for accuracy, interpretation, originality,
confidentiality, and the consequences of publication
\citep{resnikHosseini2025,icmje2026}. These responsibilities are not
removed when a GenAI system produces fluent
text or plausible analysis.
They become harder to document when the system contributes to idea
generation, literature search, coding, statistical analysis, or
peer-review reports. Disclosure can help clarify these contributions,
but it cannot by itself establish correctness, protect confidential
information, or determine whether researchers have retained
independent judgment.

Existing scholarship has begun to describe the main risks, but it has
not resolved how journals should translate those risks into workable
editorial rules. Studies of journal policies and editors' views show
substantial variation in what journals permit, what they require
authors to disclose, and how they treat reviewer use of GenAI tools
\citep{moorhouseEtAl2025,yinEtAl2025}. Several concerns recur across
this literature. First, GenAI can reproduce or amplify biases
embedded in training data, so its outputs may appear neutral while
reflecting systematic limitations in the underlying corpus
\citep{benderFriedman2018}. Second, GenAI-assisted publication
workflows
may create new paths for paper mills, fabricated references, and
authorship abuse \citep{kendallTeixeiraDaSilva2024,kendall2025}.
Third, GenAI-assisted peer review raises distinct confidentiality and
intellectual property concerns because reviewers are evaluating
manuscripts that do not belong to them
\citep{davisonEtAl2024,ebadiEtAl2025}. For statistical journals,
these concerns extend to the validity of mathematical proofs, the
design and implementation of simulations, and the reliability of
statistical inference. Journals must translate broad ethical
principles into practical decisions about which uses to permit, what
authors and reviewers should disclose, and how compliance can
reasonably be assessed.

This article reports and interprets the invited panel discussion
``Generative AI in Publishing: An Editors' Panel on Ethics and
Policies,'' held at the 2026 Joint Statistical Meetings on August 2,
2026. The session was organized by Jun Yan and Mark Glickman, chaired
and moderated by Glickman, sponsored by the Committee on Data Science
and Artificial Intelligence, and co-sponsored by the Section on
Statistical Learning and Data Science and the Section on Text
Analysis. The panelists were Paul Albert, Co-Editor-in-Chief of
\emph{Statistics in Medicine}; Yulin Hswen, Associate Editor of the
\emph{Journal of the American Medical Association} (\emph{JAMA}) and
\emph{JAMA+AI}; Lan Wang, former Co-Editor of the \emph{Annals of
Statistics}; Minge Xie, former Editor of \emph{The American Statistician}
(TAS) and current Co-Editor-in-Chief of the \emph{New England Journal of
Statistics in Data Science}; and Jun
Yan, former Editor and current Co-Editor of the \emph{Journal of Data
Science}. These roles gave the discussion a range of perspectives
from biomedical statistics, medical publishing, mathematical
statistics, data science, and general statistical scholarship. This
article is not a verbatim transcript. It is an edited and
thematically organized account, prepared from the session recording
and a speaker-labeled transcript. Attributed views are intended to
preserve the discussion while recognizing that speakers' comments
should not be read as official positions of their journals,
publishers, universities, or professional societies unless explicitly
stated.

The article contributes a comparative account of how journal editors
translate general principles governing GenAI use into practical
editorial decisions. The discussion centers on five questions: what
current journal policies permit or prohibit; what authors should
disclose; why GenAI systems cannot assume authorship responsibility;
how reviewers and editors may use GenAI without violating confidentiality
or delegating judgment; and how journals should respond to changes in
submission volume, manuscript quality, reviewer scarcity, and
research training. The panel revealed shared commitments to
transparency and human accountability alongside differing views on
their implementation. Disagreements concerned the distinction between
language assistance and idea generation, the conditions under which
author consent might permit GenAI-assisted review, and the feasibility
and desirability of enforcing restrictions on undisclosed use. By
examining these tensions, the article clarifies the choices facing
statistical journals as they seek to preserve the reliability,
fairness, and credibility of scholarly publishing.

\section{Current Journal Policies}

The journal policies described by the panelists shared a commitment
to human accountability but differed substantially in their details.
Albert explained that Statistics in Medicine, published by Wiley,
followed the publisher's policy, emphasizing that GenAI could not be an
author and that human authors remained responsible for the work. Wang
reported that, at the time of the panel, the Annals of Statistics had
no GenAI-specific policy for authors or reviewers, but its general
standards for authorship, research integrity, and peer-review
confidentiality remained applicable. Hswen described
\emph{JAMA}'s approach as more prescriptive, with disclosure 
of the use, prompt, and model and restrictions on GenAI use for
Letters and opinion writing. Xie summarized TAS and American Statistical
Association (ASA) requirements as disclosure of GenAI assistance and
the affected part of a submission.
Yan described the Journal of Data Science policy as permitting
responsible use as long as authors identify the tool and version,
explain its purpose, and validate the manuscript\textquotesingle s
accuracy, originality, and integrity.

The discussion also established that an apparent journal policy may
originate from a publisher, a society, or an individual journal.
Albert clarified that Wiley\textquotesingle s rule extended beyond
Statistics in Medicine. Wiley now specifies disclosure for GenAI that
substantially edits, develops, or translates any part of a
manuscript, while exempting basic spelling and grammar assistance
\citep{wiley2026best}. \emph{JAMA} guidance issued after the panel
specifies permitted and prohibited uses, including restrictions involving
references, clinical images, opinion articles, and responses to
reviewers \citep{flanagin2023nonhuman,flanagin2026updated}. As one
non-panel comparison, Elsevier likewise combines a publisher-wide
disclosure rule with human oversight and separate reporting of
research uses in the Methods section \citep{elsevier2026generative}.

The panel therefore identified convergence on general principles
without establishing a uniform operational policy. GenAI systems are
excluded from authorship, named authors retain responsibility, and
consequential assistance should be disclosed. The policies differ,
however, in the required level of detail and in their treatment of particular
article types and research tasks. They also leave different amounts
of discretion to journal editors. These differences are important
because a general requirement of transparency does not determine
whether authors must report routine language editing, retain prompts,
describe iterative idea development, or document assistance with
code, simulations, and proofs.

\section{Authorship Responsibility and Research Integrity}

\subsection{Accountability and Acceptable Use}

Authorship responsibility provided the panel\textquotesingle s main
criterion for evaluating GenAI assistance. The participants did not
define acceptable use by the amount of generated text or code alone.
Instead, they repeatedly asked whether the named authors understood
the work, verified its contents, supplied an identifiable
intellectual contribution, and could answer for every claim. Xie
argued that journals could neither prevent nor reliably police GenAI
use and should evaluate the resulting scholarship on its merits,
subject to transparency and human ownership of the intellectual work.
Albert similarly described GenAI as a research assistant whose output
must be understood by the researcher. This responsibility-based
approach did not make every use permissible. It established
conditions under which assistance could be considered and provided a
basis for distinguishing disclosed but deficient work from
fabrication, plagiarism, or other research-integrity violations.

The panel expressed a range of views about how far acceptable
assistance could extend into the intellectual process. Wang
recognized the benefits of GenAI assistance with language editing and
identifying relevant prior work, particularly for authors writing in
a second language, but expressed reservations about relying on GenAI
to generate research ideas or produce an entire paper. Yan favored a
broader rule that could allow assistance with writing, coding,
proofs, formulation, and idea generation because, in his experience,
ideation remained human-led within an iterative interaction. He also
reported that his co-editor preferred to limit idea generation.
Albert favored disclosure at every level, from writing assistance to
simulation studies and other intellectual uses. These positions did
not produce a shared boundary between routine and substantive
assistance. They instead located the disagreement within a shared
requirement: authors must identify their own contribution and remain
capable of defending the resulting research.

The exclusion of GenAI systems from authorship followed from
accountability rather than from a claim that these systems can never
contribute intellectually. Albert observed that a GenAI system cannot
be held responsible for a publication. Hswen connected that point to
established authorship criteria, and Wang separated intellectual
contribution from the ability to approve submission and accept
responsibility. An audience member brought the issue into sharper
focus by asking how credit should be assigned if a powerful system
produces correct and creative research but cannot be named as a
coauthor. Wang and Xie acknowledged that GenAI may eventually supply a
major share of an intellectual product, even though current
authorship rules cannot accommodate it. The discussion therefore
supported detailed contribution disclosure, but it did not resolve
future questions of credit, ownership, or liability when the human
contribution becomes comparatively limited.

Several uses of GenAI tools crossed a clearer research-integrity
boundary. Yan identified fabricated data, plagiarism, and nonexistent
references as unacceptable, noting that false citations can
contaminate later literature when other authors repeat them. Xie
described an experience in which he rejected a manuscript after
discovering a fabricated reference, and another submission whose
disclosed GenAI-generated design and simulation appeared complete but
lacked adequate depth. The latter case is analytically distinct from
fabrication: transparent GenAI use does not require acceptance when
the resulting scholarship is weak. Albert also reported seeing more
submissions whose parts did not fit together, although he could not
determine whether GenAI had caused the pattern. These examples
establish
that disclosure neither excuses false material nor substitutes for
editorial evaluation. A fluent or formally complete manuscript must
still satisfy the journal\textquotesingle s standards for
correctness, coherence, depth, and contribution.

Images and mathematical proofs illustrated why correctness requires
domain-specific verification. Hswen opposed the use of GenAI-generated
or GenAI-enhanced images of human specimens, cells, tissues, or
organisms because such alterations may misrepresent the underlying
scientific evidence. Wang and Albert likewise treated fabrication or
manipulation of research data as a firm boundary. For theoretical
research, Wang warned that a GenAI-generated proof may appear
confident
while containing an error that authors and reviewers fail to detect.
An audience member asked whether translation into a formal language
such as Lean \citep{de2021lean} could make proof checking more
reliable. Xie considered formal checking potentially useful, whereas
Albert questioned whether a GenAI checker might reproduce the
generator\textquotesingle s error. The panel did not resolve that
technical question, but it agreed that neither plausible-sounding
text nor confirmation from a second automated system should be
treated as proof of correctness.

The audience discussion of synthetic data further clarified that the
method of generation does not by itself determine whether data use is
legitimate. An audience member noted that synthetic data, images, and
maps may protect human participants and are common in medical
research. Hswen stated that research using synthetic data should
clearly disclose its use, compare it with relevant real-world data,
and validate its suitability for the intended purpose. Xie
distinguished blind substitution from established uses of simulation,
including simulation-based inference for complex models with
intractable likelihoods. From this perspective, synthetic data are
not equivalent to fabricated observations when their origin is
reported and their role is scientifically justified. The integrity
question concerns whether authors understand the generating process,
validate its consequences, and avoid presenting synthetic evidence as
direct observation, rather than simply whether the data are synthetic.

\subsection{Disclosure of GenAI Use}

Disclosure made the principle of authorship responsibility
operational by informing editors, reviewers, and readers how GenAI had
contributed to the work. The panel discussed assistance with
language, drafting, coding, simulation, literature search, idea
formulation, proofs, references, and images, but did not agree on a
single reporting rule for all uses. Hswen described a detailed,
prescriptive \emph{JAMA} approach that permits GenAI for research and
manuscript preparation when fully disclosed. Authors must identify
the tool, version, developer, date, purpose, and affected content,
verify all outputs, and accept full responsibility for their accuracy
and integrity. GenAI use is prohibited for drafting opinion articles,
letters, peer-review responses, and clinical images, while its use
for references is discouraged. Xie emphasized the affected portion of
the manuscript, and Yan emphasized the tool, version, and purpose of
use. Albert argued that authors should distinguish the human and GenAI
contributions when assistance is intellectual in nature, such as idea
generation, simulation design, mathematical formulation, or proof
development. Glickman noted that a simple statement that GenAI was
used, without specifying its purpose, gives an editor little basis for interpreting
a disclosure. Taken together, these views support a functional
disclosure that identifies what the system contributed and where that
contribution entered the research, rather than a binary declaration
detached from the manuscript\textquotesingle s development.

An example described by an audience member raised the practical limit
of highly granular disclosure. The audience member had consulted many
models through approximately one hundred interactions and could no
longer reconstruct every model or prompt, although the systems had
assisted with the model formulations. Albert recommended reporting
that broad pattern directly rather than attempting an impossible
reconstruction. Xie agreed that editors needed an account of the
process and methodological role, not an extensive prompt dossier.
Hswen emphasized that detailed reporting of the GenAI tool, version,
prompts, dates of use, and role in the research process is essential
for evaluating and reproducing the work, while Glickman observed that
trying multiple models may affect post-selection inference when
authors select an output after exploration, making it crucial to
acknowledge the modeling process. These comments suggest a
proportional standard: authors should preserve specific records when
feasible, but an honest summary of purposes, breadth, and affected
components may be more informative than incomplete prompt-level
detail for iterative work.

Disclosure nevertheless remains an incomplete mechanism for research
integrity. Glickman observed that repeated interaction can make the
boundary between an author\textquotesingle s reasoning and GenAI
assistance difficult even for an honest author to reconstruct. Xie
argued that editors should judge merit rather than attempt to detect
how much GenAI was used, and several panelists noted that prohibitions
would be difficult to police. The panel consequently relied heavily
on author honesty without identifying a reliable detection or
enforcement procedure. It also left open how journals should
distinguish minor editing from substantive intellectual assistance
and how they should respond when an accurate, disclosed paper depends
extensively on GenAI. The areas of convergence were narrower: authors
must not shift responsibility to a tool, disclosure should describe
consequential assistance, and editors must still assess the validity
and scholarly value of the submitted work.

\section{GenAI in Peer Review and Editorial Work}

GenAI raises distinct concerns in peer review because
reviewers are entrusted with confidential material that is not theirs
to share \citep{perlis2025artificial}. The panel therefore organized
this part of the discussion around confidentiality and human
editorial judgment rather than around disclosure alone. A reviewer
who transfers a manuscript to an external system may expose
unpublished text, data, methods, or ideas to a third party, such as
the company that developed the GenAI system, and may permit their
retention or later use. Hswen framed this primarily as a legal and
intellectual-property issue: authors retain rights to their writing,
and manuscripts should not be uploaded to GenAI systems without their
permission. Even when no disclosure occurs outside the software
interface, the transfer may be at odds with an author\textquotesingle
s understanding of how their manuscript will be processed. The panel
also distinguished assistance with a report from delegation of the
evaluation itself. GenAI assistance may help a reviewer express an
independently formed analysis under specified conditions, whereas
delegating the review entirely to a GenAI tool replaces the expert
judgment for which the reviewer was selected. These two concerns
defined the practical limits considered by the panel.

The represented journals varied in whether those limits appeared in
GenAI-specific rules. Albert described Wiley policy as prohibiting a
reviewer from submitting the manuscript to a GenAI system or using it
to generate a referee report, while allowing GenAI assistance that
polishes or organizes the reviewer\textquotesingle s own thoughts.
Wang explained that the Annals of Statistics relied on its general
confidentiality policy, which prohibits reviewers from sharing
manuscripts without authorization. She also cited Operations
Research, an Institute for Operations Research and the Management
Sciences (INFORMS) journal, as an example of a journal that explicitly
prohibits reviewers from uploading manuscripts to GenAI
systems that retain data or use submitted content for training. Yan
stated that the Journal of Data Science had no explicit
reviewer policy and that this omission had become an active editorial
question. Xie described the ASA guidance as the minimum for editors
and reviewers, although the ASA Publication Committee had discussed
updating the policies, especially those governing GenAI use in peer
review. The policy variation thus concerned codification and
permitted assistance, not whether reviewers retained responsibility
for their reports. Albert also considered it possible to disclose
permitted reviewer assistance, while maintaining the prohibition on
transferring or uploading the manuscript to a GenAI platform.

Panelists reported that formal restrictions had not prevented
apparently GenAI-generated reviews. Albert described receiving a report
whose format made extensive GenAI generation evident and noted the
ethical tension created because the report recommended acceptance.
Wang had similarly received reviews containing signs of GenAI use and
identified a concern beyond confidentiality: criticisms presented as
expert review may instead originate from a system. Hswen explained
that \emph{JAMA} records concerns about reviewer performance,
including 
suspected GenAI use, but she also noted the difficulty of proving a
violation and acting on it when qualified reviewers are scarce. Xie
raised the same enforcement problem for journals that depend on
volunteers. These accounts did not establish the prevalence of
GenAI-generated reviews. They demonstrated that recognizable cases
already create decisions about evidentiary standards, sanctions, and
whether the editorial recommendation can still be used.

The panel treated confidentiality as an ethical,
intellectual-property, and potentially legal obligation. Albert
compared a GenAI upload with giving a manuscript to an unapproved
colleague: both transfer privileged material beyond the authorized
review process. Hswen and Glickman added that the author may still
own the submitted text and its ideas, so a reviewer cannot
unilaterally authorize external processing. Xie consequently
discussed an author-choice model in which a submission could receive
human-only or GenAI-assisted review, although he emphasized that the
proposal remained debated and had not been implemented. Glickman
further distinguished between GenAI tools that share inputs or
incorporate them into training data and enterprise tools that keep
inputs within an institution and do not share them with the system's
developer. Wang proposed that a secure system operated by a journal
or publisher would alter the confidentiality analysis. These
proposals identify authorization and technical control as relevant
conditions, but they do not by themselves establish that GenAI
evaluation is reliable or editorially appropriate.

Even with confidentiality safeguards in place, the appropriate role
of GenAI in peer review remained contested. Albert rejected uploading
a paper and requesting a review, but he was more receptive when a
reviewer first formulated an expert assessment and then sought
assistance based on those original thoughts. Glickman suggested that
journals provide concrete guidance that would require reviewers to
read the paper, identify the central issues, and possibly use GenAI
tools to help refine the review. Xie likewise argued that GenAI could
serve only as an assistant and that reviewers must own every point in
the submitted report. Wang cautioned that differences in prompting
skill could shift, rather than reduce, variation in review quality.
Albert, somewhat facetiously, questioned whether journals would still
need nominal human reviewers if editors could obtain the same
automated report directly. Hswen questioned whether increasing
reliance on GenAI could weaken reviewers' subject-matter expertise and
their ability to critically evaluate GenAI-generated conclusions, and
Glickman identified reviewer training as a potential resulting
requirement. The discussion therefore preserved human responsibility
but did not settle which forms of reviewer-led assistance preserve
independent expert evaluation.

Editorial uses with narrower inputs offered more concrete near-term
possibilities. Yan described the Journal of Data Science
reproducibility check conducted near acceptance using supplemental
data and code. The journal had tested GenAI agents on papers already
in its publication queue, and graduate-student reproducibility checkers
found the generated reports useful. Yan also described a LaTeX
technical checker that authors could run before submission, which
would address style and formatting without transferring a
confidential manuscript through the review process. He proposed that
a GenAI review might eventually supplement human reports and that the
automated report could be shared with human reviewers, provided the
authorization problem was resolved. These examples separate
reproducibility and technical checks from decisions about novelty,
validity, and publication. The panel offered no general endorsement
of GenAI reviewers or editors.

\section{GenAI and the Research Ecosystem}

\subsection{Training Junior Researchers}

The panel emphasized the importance of preparing junior researchers
to use GenAI responsibly and ethically. Albert considered it unrealistic
to expect trainees not to use GenAI, and compared the system to a
research assistant whose output the researcher must understand. He
described cases in which postdoctoral researchers returned plausible
material from GenAI but could not explain it when questioned. Wang
reported similar experiences with doctoral students who used GenAI
proficiently but did not fully understand the methods or reasoning
behind its outputs. She stressed that students should retain
responsibility for their own reasoning and that this expectation
should be established before doctoral training. Hswen compared GenAI
with a calculator: children still learn how to do mathematics before
relying on a calculator because they must understand the reasoning,
recognize errors, and know when an answer does not make sense.
Researchers similarly need the foundational knowledge to explain,
verify, challenge, and reject GenAI-generated output, not merely the
ability to produce it.

Assessment and mentorship supplied two mechanisms for testing that
capacity. Hswen described applying this principle in her GenAI and
public health course, drawing inspiration from elementary and middle
schools that are returning to paper-and-pencil learning. She uses
oral examinations, presentations, debates, and direct questioning to
ensure that students can think independently and defend their work
without GenAI assistance. Yan similarly required students to identify
the tools and affected parts of a final project and to be prepared to
explain and defend the result. Neither proposal treated oral
performance as proof that a claim was correct; each used it to assess
whether the student had retained intellectual control. Albert, Wang,
Xie, and Yan also placed responsibility on advisors and senior
colleagues to model responsible practice, state expectations, and
identify fraud or plagiarism. Xie added that promotion standards
within a research unit may influence junior behavior more directly
than journal rules. The educational problem consequently extends from
classroom assessment to mentoring and professional evaluation.

The panel did not agree on one institutional format for GenAI ethics
education. Albert supported instruction in principle but questioned
whether the technology was too new and rapidly changing to justify
immediate curricular revision. Wang favored a course or short
orientation for incoming students. Yan proposed annual adaptive
instruction through departmental orientation and academic ethics
training. In contrast, Hswen argued that another routine online
compliance module would be insufficient. Drawing on her GenAI and
public
health course, she called for clear ethical expectations and rigorous
minimum standards across classes, including zero tolerance for
plagiarism, fabricated citations, falsified data, and
misrepresentation of GenAI-generated work, while acknowledging the
need to preserve academic freedom. Xie preferred a small number of
updatable lectures within a broader course that also addressed
intellectual property and related issues. These alternatives respond
differently to the same constraint: rules can become obsolete
quickly, yet inconsistent expectations across courses and
institutions leave students without a stable account of responsible
conduct. Recurrent instruction also responds to Yan\textquotesingle s
concern that habitual delegation to GenAI tools may gradually erode
researchers\textquotesingle{} ability to perform core scholarly tasks.

\subsection{Effects on Journals and Scholarly Communication}

The panelists associated GenAI with rising submission volume, but
their observations did not establish a common causal pattern or a
common effect on manuscript quality. Xie reported that ASA journals
had experienced substantial submission growth and considered GenAI a
probable contributor. Albert had seen more papers that appeared
incoherent or unsuitable and had become more likely to reject such
submissions quickly, although he could not determine whether GenAI was
responsible. Xie described a contrasting consequence: conscientious
authors could use GenAI to improve presentation, which made papers
more difficult to reject at initial screening and increased assignments to
associate editors. Hswen also noted that unusually long and detailed
reviewer reports may indicate the use of GenAI and can create
additional work for editors and authors. The panelists framed this
increased burden and the changing surface quality of submissions as
editorial observations, not a general conclusion that GenAI has
uniformly lowered or improved scientific quality. Hswen warned that
GenAI could inundate journals with superficially credible but
low-value submissions, making substantive scholarship increasingly
difficult to identify.

The growing volume of submissions placed additional strain on an
already limited pool of reviewers. Wang reported inviting
approximately five candidates to obtain two reviews and suggested
that associate editors conduct more initial assessment so that fewer
papers reach external review. Albert had also become more selective
at the editorial stage, while acknowledging the risk that strong work
could be rejected without specialist evaluation. Proposed incentives
included distinguished-reviewer recognition and a pilot involving
reviewer payment discussed by Xie. Xie also noted that one
editorial-board member had completed twelve reviews despite an
expectation of approximately four to six. Hswen argued that the
professional culture of peer review had weakened and that reviewing
must again be taught as a core scholarly responsibility. She
cautioned that overwhelmed editors may increasingly rely on familiar
authors and established networks, resulting in the same voices being
published repeatedly while early-career researchers and new ideas are
overlooked. These responses address limited human attention through
triage, incentives, and training, but the panel supplied no evidence
that any of the suggested approaches would resolve the shortage.

Access to GenAI may broaden participation while also creating new
inequalities. Wang emphasized that GenAI language assistance can help
authors for whom English is not their first language communicate
their ideas more effectively. Albert, Wang, and Yan also anticipated
productivity gains when GenAI reduces the labor required for simulation, formatting,
reference checking, and other routine tasks, thereby allowing
researchers to devote more time to scientific questions. An audience
member raised the opposing distributional concern that the high cost
of advanced GenAI systems could create disparities in access. Glickman
and Albert observed that researchers at well-resourced institutions
may receive enterprise access while independent researchers rely on
less capable free versions. Yan extended this concern to researchers
in developing countries and linked unequal access to future
differences in productivity and in the geographic distribution of research
activity. The panel thus
described accessibility and inequality as simultaneous consequences
whose net effect cannot be inferred from improved language support
alone.

\section{Future of Journals}

The panel envisioned several possible institutional futures but did
not agree on which was most likely. One scenario would differentiate
publication channels by the type of GenAI involvement. Hswen proposed
creating a journal or subjournal devoted to work produced entirely by
humans, alongside outlets that allow programming or language
assistance, without treating one category as superior to another.
Wang noted AiraXiv, an open-access preprint platform that explicitly
welcomes both human- and GenAI-generated papers, as a possible model
for a voluntary experiment \citep{pan2026airaxiv}. An audience member
suggested requiring authors to discuss and defend their papers
through public discussion, so that publication would reflect not only
the written work but also the author\textquotesingle s understanding
of it. Hswen and Albert considered the idea promising, but Glickman
cautioned that it could disadvantage researchers who are less
confident public speakers. A public defense might therefore offer
additional evidence of authorship and understanding, but it would
introduce its own potential biases.

A more extensive scenario would permit journals focused on
GenAI-generated research. An audience member asked whether future
GenAI
systems might perform most ideation and eventually produce work that
competes with or exceeds work in human journals. Yan supported an
experimental journal without restrictions on GenAI use, reasoning that
GenAI reviewers could then participate openly in improving
submissions.
The ensuing discussion suggested that such an outlet might function
less as a conventional authorship system and more as a curation
system whose central question is whether a GenAI system's output adds
to knowledge. This model would not eliminate the unresolved issues
identified earlier in the panel, however. It would instead require
explicit rules for selection, credit, responsibility, intellectual
property, and the status of human contributors. The panel did not
determine whether an unrestricted outlet should coexist with
conventional journals or whether readers would treat its publications
as equivalent evidence.

A third scenario would integrate secure GenAI agents into journals'
peer-review system. Xie discussed a two-track model in which authors
choose between human-only and GenAI-assisted review. Glickman observed
that enterprise GenAI systems that keep submitted material within a
secure local environment might offer an alternative to publicly
available systems. Wang proposed a journal- or publisher-operated
agent and pointed to the GenAI-assisted peer-review pilot at the 40th
Annual AAAI Conference on Artificial Intelligence (AAAI-26) as a
related example. The pilot supplemented human reviews with clearly
labeled GenAI-generated assessments that contained no scores or
acceptance recommendations, while retaining human reviewers and human
authority over acceptance decisions
\citep{association2025instructions}. Albert raised a broader
question: if editors can generate an automated report themselves,
what value is added by a nominal reviewer who simply submits the same
output? Glickman further noted that authors could repeatedly obtain
self-generated automated reviews and revise their papers before
submission until the systems identified no remaining weaknesses.
Journals would therefore need to specify whether GenAI is intended to
check, supplement, or replace human review and to evaluate each use
separately for confidentiality, accuracy, bias, and editorial
responsibility.

Autonomous research agents could amplify these challenges. Wang
described systems capable of taking a data set, formulating research
ideas, and drafting much of a manuscript. Their widespread use could
increase the volume of submissions that appear complete while making
the extent of human intellectual contribution harder to establish.
The panelists also differed about whether future GenAI use would
remain detectable. Hswen anticipated that journals would develop more
effective tools and processes for identifying undisclosed GenAI use in
submitted manuscripts, while Albert anticipated that GenAI-generated
text would become increasingly harder to identify as technology
improved. This disagreement highlighted the limitations of policies
that depend on detection rather than disclosure and evaluation. It
also reinforced the case for assessing a submission by its validity
and contribution without assuming that its production process can be
reliably inferred from the final text. In a lighthearted aside, Yan
floated the idea of publication quotas, adding that researchers who
had reached their limit might collaborate with colleagues who still
had room under theirs.

\section{Conclusion}

The panel discussion highlighted the challenge of preserving
accountability as GenAI becomes increasingly integrated into research
and publishing. Despite differences on specific issues, several
shared principles emerged: GenAI systems cannot assume the
responsibilities of authorship, human authors and reviewers remain
accountable for their work, substantive GenAI assistance should be
disclosed, and confidential submissions should not be shared with
unauthorized systems. These principles do not resolve the operational
questions that occupied much of the panel. Panelists differed over
assistance with ideas and proofs, the appropriate detail of
disclosure, permissible reviewer support, and the institutional role
of automated evaluation. The discussion favored policies that clearly
define responsibilities and protect confidential information while allowing
editors to exercise judgment about contested use of GenAI and revise
those policies as the technology evolves.

The practical priority is to evaluate new policies and editorial
practices rather than assume that either prohibiting or adopting
GenAI will have the intended effects. Journals, publishers,
professional societies, and research institutions need coordinated
guidance for authors, reviewers, editors, and trainees, along with
procedures that can evolve as tools and secure infrastructures
change. Evidence is needed on disclosure burdens, submission volume
and quality, editorial decision-making, the speed and accuracy of
peer review, disparities in access, and research-integrity failures.
New publication formats, automated checks, hybrid review models, and
submission limits should be assessed against these outcomes before
they are widely adopted. Although the panel did not produce a unified
set of recommendations, it clarified the choices confronting
statistical publishing and the evidence needed to guide them.
Addressing these questions will require policies that combine clear
accountability with a willingness to learn from experience and revise
practices as the technology develops.

\section*{Acknowledgments}

The authors thank the audience participants whose questions and
comments extended the panel discussion. The session was sponsored by
the ASA Committee on Data Science and
Artificial Intelligence and co-sponsored by the Section on
Statistical Learning and Data Science and the Section on Text
Analysis.

\section*{Data Availability Statement}
No research datasets were generated or analyzed for this article.

\section*{Disclosure of GenAI Use}

An automated iPhone transcription provided one source transcript,
which was reconciled with the session recording and edited to produce
the speaker-labeled transcript. OpenAI Codex, accessed in August
2026, assisted with thematic organization, manuscript drafting and
revision, consistency checking, and review of current public policy
sources. Attributed remarks were checked against the transcript. The
authors retain responsibility for the accuracy, originality,
interpretation, and integrity of the article.

\bibliographystyle{asa}
\bibliography{refs}

\end{document}